\documentclass[12pt]{elsart}
\usepackage{epsfig}

\begin{document}

\begin{frontmatter}

\title{Multifragmentation process for different mass asymmetry in the 
entrance channel around the Fermi energy}

\author[1]{N.~Bellaize}
\author[1]{O.~Lopez} 
\author[2]{J.P.~Wieleczko}
\author[1]{D.~Cussol}
\author[2]{G.~Auger}
\author[3]{Ch.O.~Bacri}
\author[1]{F.~Bocage}
\author[3]{B.~Borderie}
\author[1]{R.~Bougault}
\author[2]{B.~Bouriquet}
\author[1]{R.~Brou}
\author[4]{P.~Buchet}
\author[1]{A.M.~Buta}
\author[4]{J.L.~Charvet}
\author[2]{A.~Chbihi}
\author[1]{J.~Colin} 
\author[4]{R.~Dayras}
\author[5]{N.~De Cesare}
\author[6]{A.~Demeyer}
\author[4]{D.~Dor\'e}
\author[1]{D.~Durand}
\author[2]{J.D.~Frankland}
\author[3,7]{E.~Galichet}
\author[1]{E.~Genouin-Duhamel}
\author[6]{E.~Gerlic}
\author[2]{B.~Guiot}
\author[6]{D.~Guinet}
\author[2]{S.~Hudan}
\author[3,8]{G.~Lanzalone}
\author[6]{P.~Lautesse}
\author[3,4]{F.~Lavaud}
\author[2]{J.L.~Laville}
\author[1]{J.F.~Lecolley}
\author[4]{R.~Legrain\thanksref{d}}
\author[2]{N.~Le Neindre}
\author[1]{L.~Manduci}
\author[1]{J.~Marie}
\author[4]{L.~Nalpas}
\author[1]{J.~Normand}
\author[3,9]{M.~P\^arlog}
\author[3]{P.~Paw{\l}owski} 
\author[3]{E.~Plagnol}
\author[3]{M.F.~Rivet}
\author[5]{E.~Rosato}
\author[10]{R.~Roy}
\author[2,11]{F.~Saint-Laurent}
\author[1]{J.C.~Steckmeyer} 
\author[3,9]{G.~T\u{a}b\u{a}caru}
\author[1]{B.~Tamain}
\author[1]{E.~van Lauwe}
\author[3]{L.~Tassan-Got}
\author[1]{E.~Vient}
\author[5]{M.~Vigilante}
\author[4]{C.~Volant}

\address[1]{Laboratoire de Physique Corpusculaire, IN2P3-CNRS, 
ISMRA et Universit\'e, F-14050 Caen Cedex, France}
\address[2]{Grand Acc\'el\'erateur National d'Ions Lourds, 
CEA et IN2P3-CNRS, B.P.~5027, F-14076 Caen Cedex, France}
\address[3]{Institut de Physique Nucl\'eaire, IN2P3-CNRS, 
F-91406 Orsay Cedex, France}
\address[4]{DAPNIA/SPhN, CEA/Saclay, F-91191 Gif sur Yvette Cedex, France}
\address[5]{Dipartimento di Scienze Fisiche e Sezione INFN, 
Universit\`a di Napoli ``Federico~II'', I-80126 Napoli, Italy}
\address[6]{Institut de Physique Nucl\'eaire, IN2P3-CNRS et Universit\'e,
F-69622 Villeurbanne Cedex, France}
\address[7]{Conservatoire National des Arts et M\'etiers, 
F-75141 Paris cedex 03, France}
\address[8]{Laboratorio Nazionale del Sud, Via S. Sofia 44, I-95123, 
Catania, Italy}
\address[9]{National Institute for Physics and Nuclear Engineering, 
RO-76900 Bucharest-M\u{a}gurele, Romania}
\address[10]{Laboratoire de Physique Nucl\'eaire, Universit\'e Laval, 
Qu\'ebec, Canada}
\address[11]{DRFC/STEP, CEA/Cadarache, F-13018
Saint-Paul-lez-Durance Cedex, France}
\thanks[d]{deceased}

\vskip 1cm

\begin{abstract}
The influence of the entrance channel mass asymmetry upon the fragmentation process is 
addressed by studying heavy-ion induced reactions around the Fermi energy. The data have 
been recorded with the INDRA $4\pi$ array. An event selection method 
called the Principal Component Analysis is presented and discussed. It is applied 
for the selection of central events and furthermore to multifragmentation of
single source events. The selected subsets of data are compared to the 
Statistical Multifragmentation Model (SMM) to check the equilibrium
hypothesis and get the source characteristics. Experimental comparisons show the 
evidence of a decoupling between thermal and compressional (radial flow) component 
of the excitation energy stored in such nuclear systems. 
\end{abstract}


\end{frontmatter}

\section{Introduction}

The characterization of the equation of state of nuclear matter can be
carried out by the study of the fragmentation process in heavy-ion induced
reactions \cite{bertsch,schulz,bonche}. In the Fermi energy range, the interplay between 
the attractive
mean-field interaction and the individual nucleon-nucleon scattering is
reached and the time scales of the collision become comparable
to the deexcitation times \cite{suraud}. In this case, one could expect to 
observe some deviations to the simplistic scenario consisting in the 
formation and de-excitation of a hot
and equilibrated nucleus and the role of the dynamics of the collision could
then be probed and quantified. The multifragmentation process, 
occuring in very dissipative reactions and successfully described as a 
statistical process \cite{gross}, can be studied in order to shed light on this issue.

In this work, we have isolated and studied the characteristics of
multifragmentation samples of central events for the $^{58}$Ni + $^{197}$Au
system, ranging from $32A$ to $90A$MeV, and for the 
$^{129}Xe+^{nat}Sn$ system at $50A$ MeV\footnote{Experiments performed at GANIL}, 
both registered with the 
INDRA $4\pi$ array. Both theoretical and
experimental comparisons were performed and have allowed to address
the issue of the fragmentation process and its relationship to the dynamics
of the collision. The article is organised as follows. The second section presents the
experimental details about the studied systems. The third section 
is devoted to the presentation of a new event
selection method of the most violent collisions (called 
{\it central collisions} hereafter for the sake of simplicity). This method 
will allow us to carefully isolate in a second stage multifragmentation event samples. 
The gross features of the selected events is presented in the fourth section 
and the comparison protocol is detailed.
The fifth section
presents the comparison of the selected samples with a statistical
model (SMM) in order to evaluate their degree of equilibration and will discuss
in what extent a statistical approach can account for the deexcitation process.
At last, the sixth section compares systems with different entrance channel
characteristics (mass asymmetry and incident energy), in order to give an 
experimental answer to the question of the influence of the dynamics 
upon the deexcitation process of hot composite systems formed in heavy-ion 
induced reactions in the Fermi energy domain.

\section{Experimental details}

The Nickel beam 
was accelerated through the two coupled-cyclotrons at $74A$ and $90A$ MeV 
and then degraded in order to cover the incident 
energy range between $32A$ and $90A$ MeV. The beam intensity 
was below $10^8$ ion/s in order to keep the random coincidences as low as 
possible ($< 10^{-4}$).
 The projectiles were impiged on a $200\mu g/cm^2$ self-supported $Au$ target. 
 The experimental setup was constituted by the 
 INDRA $4\pi$ array \cite{pouthas}. It is composed by $336$ two- or three-layers
 telescopes, depending on the polar angle, disposed in a cylindrical geometry 
 along the beam direction into $17$ rings from $2$ to $176^o$ covering $90\%$
 of $4\pi$. The first ring 
 ($2<\theta<3^o$) was made of $12$ phoswich plastic scintillators ($NE115$/$NE102$).
 The forward part (from $3$ to $45^o$) is constituted by 180 modules ($48$ ionisation chambers 
 filled with $C_3F_8$ gas at low pressure, $180$ silicon 
 detectors ($300\mu m$-thick) coupled to CsI(Tl) scintillators read out 
 by photomultipliers). 
 The backward part (above $45^o$) is formed by telescopes made of $48$ 
 ionization chambers associated to $144$ CsI(Tl) scintillators. The
energy range extends from $1A$ MeV to $250A$ MeV and corresponds 
to a large dynamics with a good energy resolution thanks to the $2$ 
simultaneous amplification levels implemented in the $12$ bits-QDC modules for the 
ionization chambers and the silicon diodes 
(about $70$ / $1000$ keV per channel respectively). Isotope identification is 
achieved for light charged particles 
(from $Hydrogen$ up to $Beryllium$ isotopes) for energies larger than 
$6A$ MeV. Charge identification is
obtained over the whole energy range from $Z=1$ up to $Z=79$ with a
resolution better than one charge unit for the forward part while it overcomes 
slightly this value for $Z>20$ at backward angles ($1-3$ units). The 
overall energy resolution has been estimated to be better than $5\%$, 
depending on the particle and the polar angle \cite{tabacaru,parlog}. 
Events were registered
 when at least four telescopes ($M>=4$) were fired and the overall acquisition dead time was kept below $20\%$. More details 
 about the performances of INDRA and its related electronics can be 
 found in Refs. \cite{pouthas,steckmeyer}. 
     
\section{Principal Component Analysis}

Numerous event selection methods have been already used in order to
select multifragmentation events 
\cite{cugnon,bougault88,phair,louvel,lopez93,marie,salou,lukasik,hamilton,magda,dagostino,gharib,kreutz,schuttauf,frankland}. 
These methods are mostly based on cuts in global variables distributions. 
One of the most difficult task is here
to select high excitation energy events (where multifragmentation is
supposed to be the dominant de-excitation process) associated with low
reaction cross-sections. Moreover, a complete study of the multifragmentation 
process needs to achieve exclusive measurements with a $4\pi $ array and 
so to deal with well-detected events, where all or almost all the reaction 
products have been measured. These conditions make the event selection one of 
the crucial part of the analysis in heavy-ion induced reactions and this section
is exclusively devoted to this issue.

\subsection{Standard event selection}

\begin{figure}[!h] 
\centering
\epsfig{file=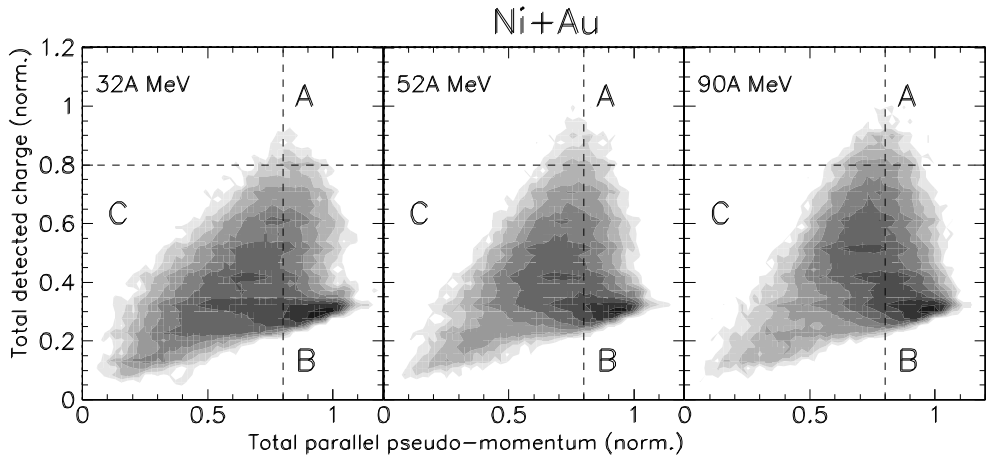,width=15cm}
\caption{$Ni+Au$ INDRA data. Correlation between the total detected charge and
the total parallel pseudo-momentum for $32A$ (left), $52A$ (middle) and $90A$
MeV (right). The dashed lines correspond to $80\%$ of the total detected charge
and parallel pseudo-momentum.}
\label{article1}
\end{figure}

For the $Ni+Au$ system, the usual methods based on the selection of 
well-detected events are not applicable because of their very small number; 
indeed, for this asymmetrical system, central events are not fully 
measured due to the energy thresholds of the experimental apparatus. 
This is illustrated by Fig. \ref{article1} for the $3$ incident energies. 
It displays the correlation between the total detected
charge $Z_{tot}$ and the total parallel pseudo-momentum $P_z^{tot}$ 
(defined as : $P_{z}^{tot} = \sum_{i=1}^{N} Z_i V_i^{//}$ where $Z_i$ and $V_i^{//}$ are 
respectively the atomic number and the parallel velocity with respect to the beam axis of the 
$i^{th}$ particule), both normalized to the
incident values. The dashed lines represent the requirement of more than 
$80\%$ of the collected charge and pseudo-momentum usually used by the 
INDRA collaboration to define complete events (zone A of Fig. \ref{article1}).

The criterion applied for this
system retains indeed too few events (less than $0.1\%$ \cite{bellaize}) 
for being reasonably considered. Obviously, we can slightly release the 
condition on the total detected charge to increase the number of retained 
events, but it turns to be somewhat arbitrary since no singularities 
appears in the $Z_{tot}$ distribution. Other selections have been also 
considered (light charged particles multiplicity or tranverse energy) but the conclusions 
remain unchanged.

Therefore, we have prefered to develop a new method which takes 
advantage of the exclusive measurements by requiring the overall information 
given by the experimental apparatus, without asking {\it a priori} any criterion 
for the detection completeness. This method is based upon the 
features of multidimensional analyses. Many kinds of multidimensional 
analyses can be employed and some of them have been already used successfully 
in the heavy-ion induced reactions domain \cite{desesquelles,maskay,bouriquet}. 
Among them, we have chosen the Principal Component Analysis ($PCA$) for 
its simplicity, and because it is closely related to more conventional 
selection methods using global variables.

\subsection{Basics}

We start from a set of global variables, commonly used in heavy-ion
physics such as multiplicities (light charged particles - LCP, $0<Z<3$ - and intermediate
mass fragments - IMF, $Z>2$ - ), transverse energy, event shape in momentum space 
(eigenvalues of the kinetic flow tensor, flow angle), detection observables 
(total detected charge, parallel momentum, etc ... ). The total number of global 
variables used in this study is $M=24$ and has been set in order to cover 
all the experimental information that can be accessed from the data after different tries. 
We then build the covariance matrix $V$ of these global variables which matrix elements 
$V_{ij}$ are defined as~:

\begin{equation}
V_{ij} = \Sigma_{k=1}^{N} \hat X_{ik} \hat X_{jk}
\label{covariance}
\end{equation}

where the indexes $i$ and $j$ run over the number of global variables ($M$) 
(squared $M \times M$ matrix), $k$ is the event number, $N$ is the total number of events 
and $\hat X_{ik}$ is the reduced value of the $i^{th}$ global variable for the event $k$. 
$\hat X_{ik}$ is built upon the original value $X_{ik}$ by applying the following linear 
transformation in order to get centred and reduced values :

\begin{equation}
\hat X_{ik} = \frac{X_{ik} - \bar X_i}{\sigma_i}
\label{reduction}
\end{equation}

where $X_{ik}$ is the value of the $i^{th}$ global variable for event $k$, $\bar X_i$ its mean
value and $\sigma_i$ its standard deviation over the total number of events $N$.

 By diagonalizing the covariance matrix $V$, we obtain the eigenvalues and 
 the associated eigenvectors, defining then {\it linear combinations} of 
 these variables, called hereafter principal components \cite{acp}. 
 These are ordered and define new planes - principal planes - where the data 
are projected. Since we use global variables, an event is then represented by only 
{\it one} point in this $M$-dimensional space. 

\subsection{Application to the data}

\begin{figure}[!h] 
\centering
\epsfig{file=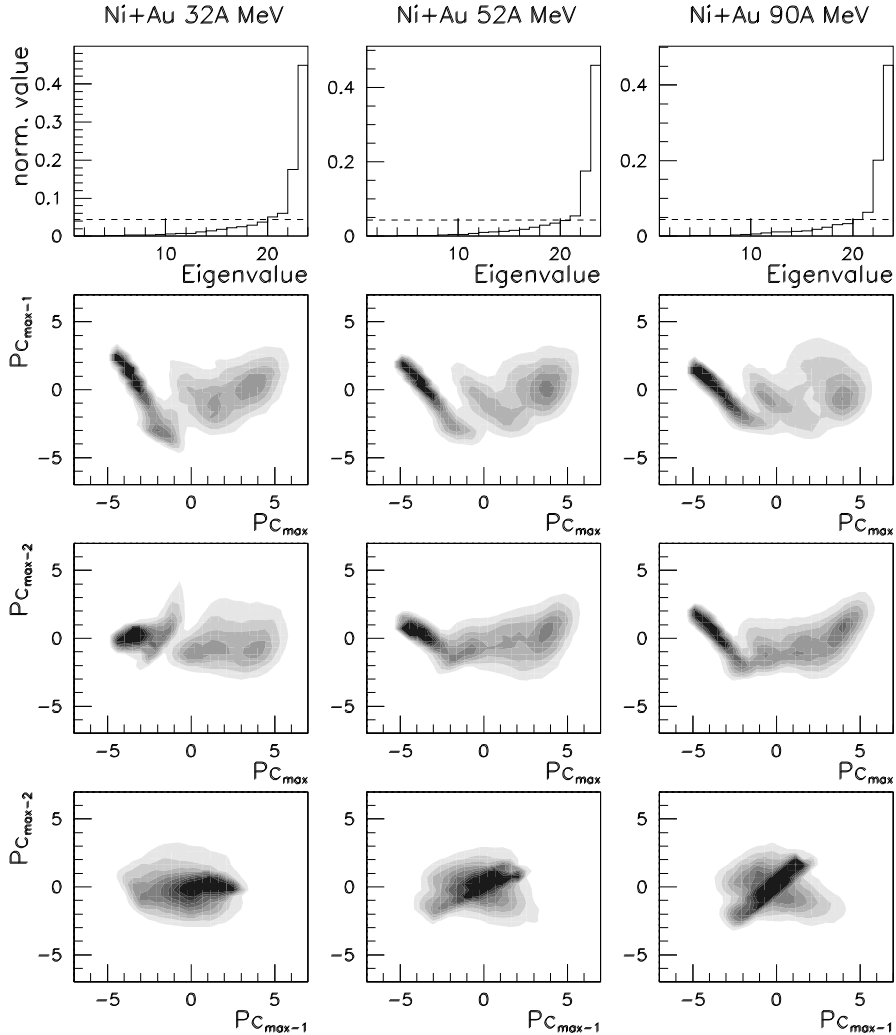,width=14cm}
\caption{$Ni+Au$ INDRA data. Top row~: Eigenvalues spectra extracted from the 
PCA for the 3 energies ($32A$, $52A$ and $90A$ MeV respectively from left to right). The
dashed line represents the threshold (here $1/24$) given by assuming a random 
information carried by each eigenvalue. Second row~: data projection in the plane defined 
by the two highest principal components ($Pc_{max} \otimes Pc_{max-1}$). Third row~: data 
projection in the plane defined by the third 
($PC_{max-2}$) and first ($PC_{max}$) principal components. Bottom row~: 
projection in the plane defined by the second ($Pc_{max-1}$) and third 
($Pc_{max-2}$) principal components.}
\label{article2}
\end{figure}

The normalized eigenvalues give the statistical information carried by the associated 
eigenvectors and the combination of the highest eigenvalues define principal planes 
where the data 
exhaust the maximum of information \cite{acp}. Fig. \ref{article2}, top row gives the
spectra obtained by diagonalization of the covariance matrix for the $Ni+Au$ data 
at $32A$, $52A$ and $90A$ MeV. Typically, we get better than $2/3$ of the total 
statistical information, called {\it explained inertia}, by projecting in the first 
principal plane ($Pc_{max} \otimes Pc_{max-1}$), which means that we have access to 
a fairly large amount of information compared to simple mono- or even bi-dimensional plots. 

Therefore, we expect to gain selectivity, compared to usual monodimensional 
event selection methods. Moreover, the $PCA$ gathers in a natural way 
events with the same features (i. e. same values of the global variables), 
defining thus naturally {\it event classes} as it is clearly seen on 
Fig. \ref{article2}. The three bottom rows of Fig. \ref{article2} displays 
the projection obtained by combining the planes defined with the 3 highest 
principal components. We observe that the second row ($Pc_{max} \otimes Pc_{max-1}$) shows
the best separation between the $4$ different event classes, as compared to the other 
planes, except for the $32A$ MeV $Ni+Au$ system (left panel). This correlates
the information brought by the normalized eigenvalues and shows that the
selection in the first principal plane is sufficient enough at this level to disentangle
between the observed event classes.

\begin{figure}[!h] 
\centering
\epsfig{file=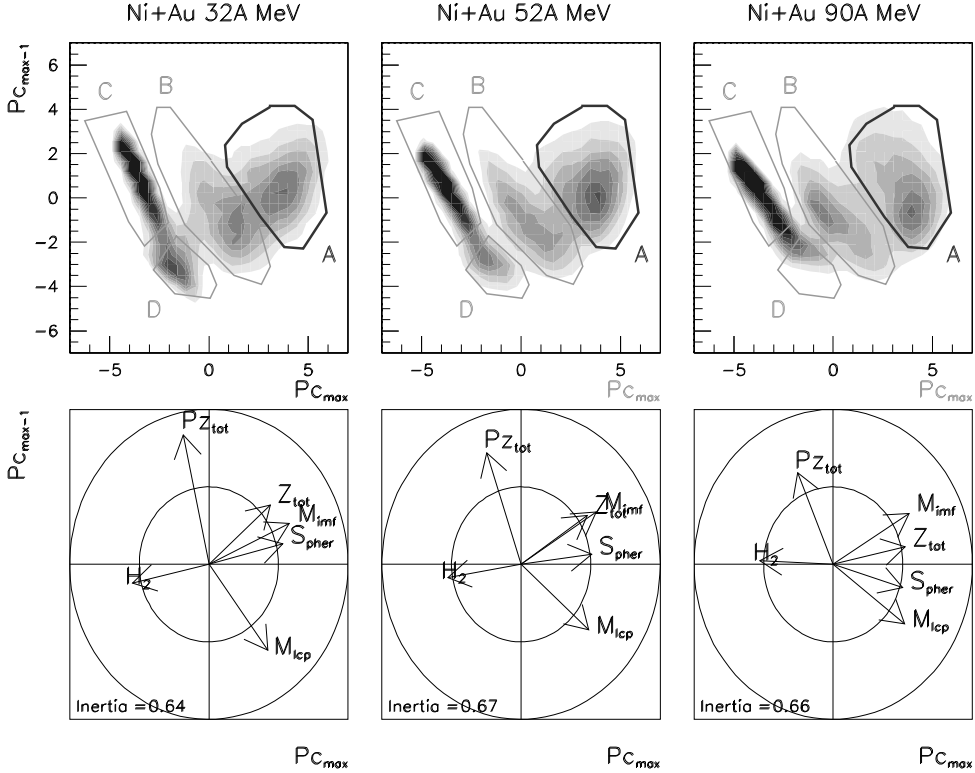,width=15cm}
\protect\caption{$Ni+Au$ INDRA data. Top row~: 
Projection in the principal plane $Pc_{max} \otimes Pc_{max-1}$ for $32A$ (left), $52A$ 
(middle) and $90A$ (right) MeV. Bottom row~: correlation plot for selected global 
variables.} 
\label{article3}
\end{figure}

Figure \ref{article3}, top row shows again the projection of the $Ni+Au$ data on
the principal plane. The inertia values, defined as the sum of the two
eigenvalues, are also reported and give the amount of information carried by the projection 
in this plane; as already mentioned, we get
around $2/3$ of the total available information. The relationship between the 
principal components and the global variables are given on Fig. \ref{article3} (bottom
row) by the so-called {\it correlation plot}. It explains in an easy way how the data are
shared out in the principal plane. For example, by looking at the total detected 
momentum along the beam axis ($Pz_{tot}$), we see that this quantity gets higher 
when we follow the direction pointed by the associated vector on the correlation plot (upper
left). The corresponding direction gives then the location of the well-detected events in
momentum; they corresponds to the most peripheral events (upper-left branch called $C$ 
on Fig. \ref{article3}) and are associated with low values of the total detected charge $Z_{tot}$ since the
corresponding two axes are perpendicular. 
Conversely, the high-multiplicity events are located almost in the opposite direction (see $M_{lcp}$ on
Fig. \ref{article3}) and are then practically anticorrelated with the total detected charge.  
The complete events (defined as the event class with
the higher values of the total detected charge $Z_{tot}$) are clearly located at the
rightmost part of the plot on Fig. \ref{article3} (called zone $A$). By
selecting the different zones as presented on Fig. \ref{article3}, we are then 
able to naturally separate and study the different event classes from peripheral 
(incomplete events) to the most dissipative collisions (pseudo-complete events).

We have checked that the
rightmost zone, called $A$, is here associated to the highest values of the
total detected charge and light charged particle multiplicity while the 
zones $B$, $C$ and $D$ are associated to smaller values as shown in 
table \ref{tabzone} for the $52A$ MeV $Ni+Au$ system. Table \ref{tabzone} gives 
the mean multiplicities (light charged 
particles and intermediate mass fragment) and the total detected charge per event. 
As already observed before from the correlation plot, 
zone $A$ corresponds to the
 highest values (the mean IMF multiplicity is equal to $4.0$), and groups the
 best-detected (total charge) and most dissipative (high multiplicity) events. In the
 following, we will only retain this zone because it contains
  the multifragmentation events (among other dissipative events) coming from
 a composite system formed in very central collisions. 
 
\begin{table}[!h]
\begin{center}
\begin{tabular}{lrrrr}\hline
  & Zone $A$ & Zone $B$ & Zone $C$ & Zone $D$   \\ \cline{1-5}  
$<M_{LCP}>$ ($Z\leq2$) & 16.6 & 13.6 & 5.0 & 14.2 \\
$<M_{IMF}>$ ($Z\geq3$)& 4.0 & 2.5 & 1.5 & 1.5  \\ 
$<Z_{TOT}>$ & 48.0 & 34.7 & 28.0 & 27.6   \\ \cline{1-5}
\end{tabular}
\vskip 0.2cm
\caption{ $52A$ MeV $Ni+Au$ data. Mean light charged particles 
multiplicity ($M_{LCP}$), IMF multiplicity ($M_{IMF}$), 
total detected charge ($Z_{TOT}$) for events 
from Zones $A$ to $D$ of Fig. \ref{article3}.}
\label{tabzone}
\end{center}
\end{table}

 Furthermore, a detailed 
 study of the other classes \cite{bellaize} -not presented here- has shown that 
 Zone $B$ is associated to incomplete fusion events where the fused system 
 undergoes fission (mid-peripheral collisions), Zone $C$ to peripheral collisions 
 where the quasi-projectile is close to the projectile (Nickel) and Zone $D$ 
 to binary dissipative collisions where a (damped) quasi-projectile still survives 
 in the exit channel.

\subsection{Multifragmentation selection}

The events belonging to zone $A$ of Fig. \ref{article3} 
(best detected and dissipative events) are kept and a {\it second} $PCA$ is 
applied on this sample in order to refine the selection on central events. 
This second $PCA$ allows a more precise ordering of the events of zone $A$ 
by enlarging the selectivity. In this way, we only need to work
on the first principal plane ($Pc_{max} \otimes Pc_{max-1}$), and we do not 
need to look at the other principal planes. 

\begin{figure}[!h] 
\centering
\epsfig{file=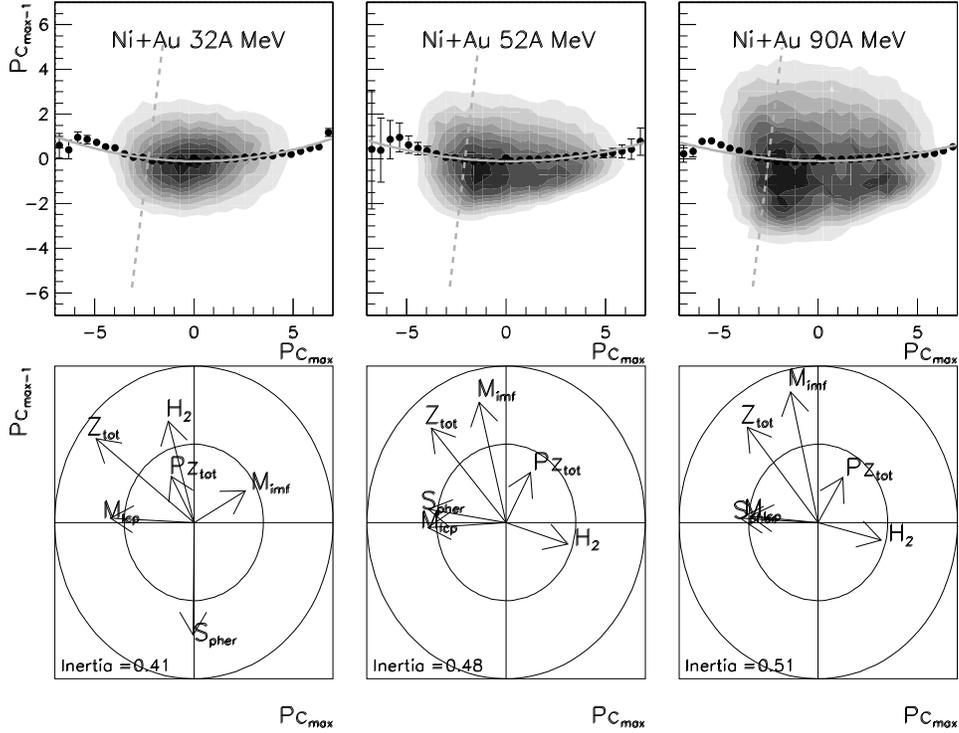,width=15cm}
\caption{$Ni+Au$ INDRA data. Second $PCA$. Top row~: 
Projection in the principal plane $Pc_{max} \otimes Pc_{max-1}$ for $32A$ (left), $52A$ 
(middle) and $90A$ (right) MeV. Bottom row~: correlation plot for selected global 
variables. }
\label{article4}
\end{figure}

The result of this second $PCA$ is displayed on Fig. \ref{article4} (top row) and the 
corresponding correlation plot is given in the bottom row. We see that the events are
grouped together in an elongated shape along $Pc_{max}$ and there is no more any
separation between event classes, showing thus that the primary selection (first $PCA$)
was sufficient to get the gross features of this selected event sample. Nevertheless,
we can define a ridge line from the mean values represented by the symbols 
(displayed as the thick curve on Fig. \ref{article4}, top row ) and then 
cut  perpendicularly as represented by the vertical dashed line on the left 
of the panels of Fig. \ref{article4}. If we now look at the
correlation plots, we see that the leftmost zone is associated to the best detected
events ($Z_{tot}$), the highest LCP multiplicity $M_{lcp}$ and the most compact shape in
momentum space (represented by the sphericity $S_{pher}$). At the opposite, the rightmost
zone is characterized by events with an elongated shape in momentum space, as represented
by the highest $H_2$ values \cite{marie}. It is interesting to note that the maximum
IMF multiplicity ($M_{imf}$) is obtained for events on the right part of the projection
plot for the $32A$ MeV $Ni+Au$ system and is associated to peripheral (elongated) and 
not central events. This surprising result is very well explained by the detection
response of INDRA; in the $32A$ MeV case, the small value of the recoil velocity of the 
composite system in central collisions does not allow to overcome the energy thresholds
of INDRA for the heavy products, so the IMF multiplicity does not reflect the true
multiplicity for the central collisions and the experimental apparatus miss more likely 
the heavy and slow fragments in this case.

\begin{figure}[!h] 
\centering
\epsfig{file=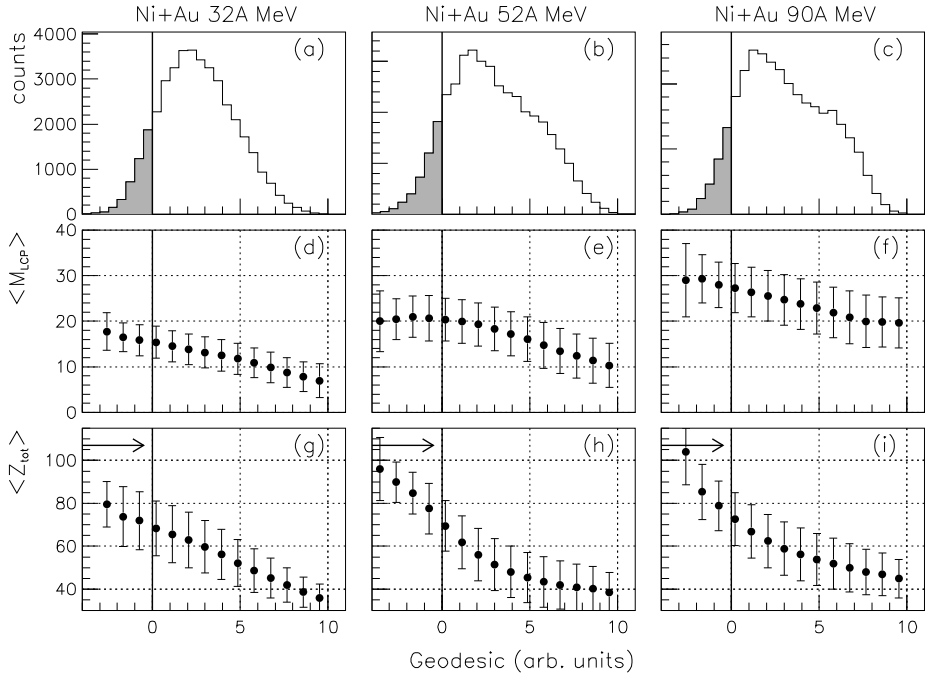,width=14cm}
\caption{$Ni+Au$ INDRA data. Geodesic (top row) distribution, Mean LCP (middle) and total
detected charge (bottom) as a function of the geodesic. The symbols correspond
to the data. The hatched area on the geodesic distributions shows the 
selected events (see text).}
\label{article5}
\end{figure}

The ridge line displayed on Fig. \ref{article4} defines a new curvilinear abscissa
called {\it geodesic} $g$ and allows a refined event sorting. Figure 
\ref{article5} presents the evolution of the geodesic (top row), 
the LCP multiplicity (middle row) and the total detected charge (bottom row) 
as a function of this curvilinear abscissa which is dimensionless : the origin of 
the curve is set at the location of the dashed line on Fig. \ref{article4}, top row 
so that events with negative $g$ values correspond to the selection 
we are going to apply on the data. The number of retained events has been chosen in order to
get the same amount of data for the different incident energies and to keep homogeneous
values for the global variables (multiplicities, tranverse energy). It corresponds to a
measured cross-section of $100mb$ ($b<1.7fm$). 

By selecting events with negative $g$ values as presented on Fig. \ref{article5}, top row 
by the hatched area, we get the highest LCP multiplicities and the best charge
detection (around and above $80\%$ the initial total charge). The selection 
operated by the cut along the geodesic retains only events 
with the highest multiplicities, compacity and detection efficiency, providing thus 
multifragmentation event samples which are extensively studied in 
the following sections.

\subsection{Generic features}

\begin{figure}[!h] 
\centering
\epsfig{file=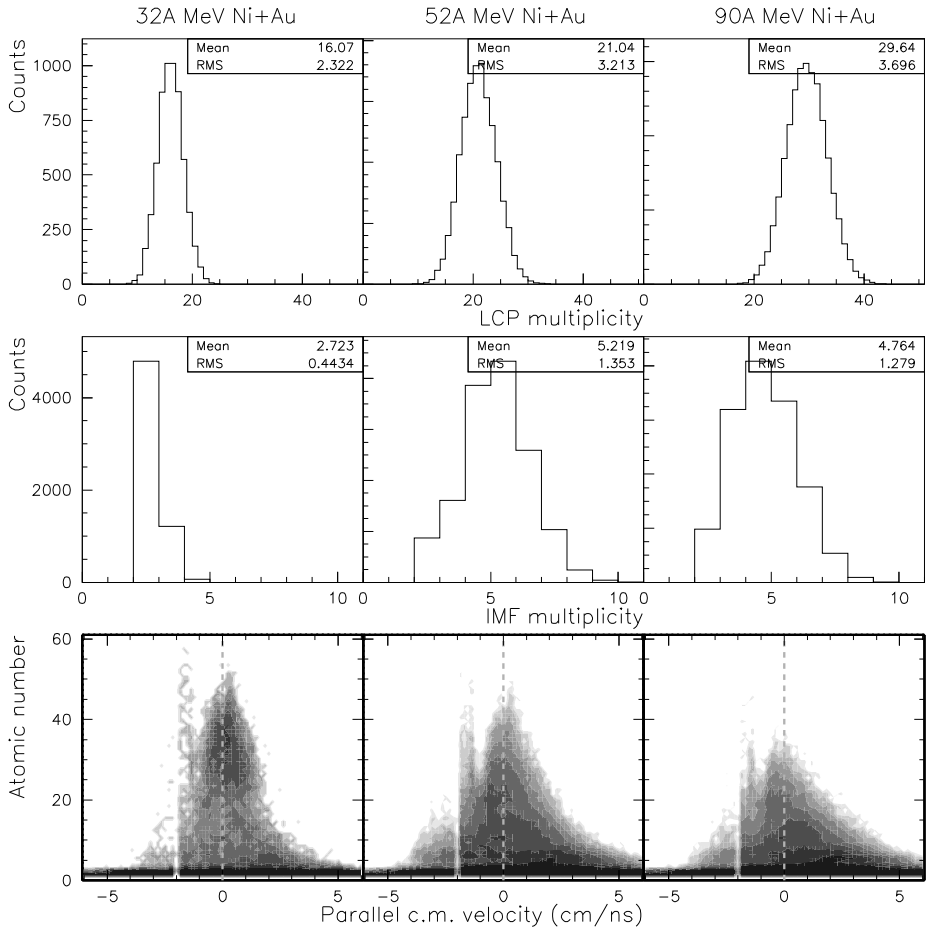,width=14cm}
\caption{$Ni+Au$ multifragmentation data. 
Top row~: LCP multiplicity.  Middle row~: IMF multiplicity. Bottom row~: 
atomic number vs parallel c.m. velocity.}
\label{article6}
\end{figure}

Figure \ref{article6} displays the LCP and IMF multiplicities, and the
atomic number vs. parallel c.m. velocity plot (along the beam direction). 
The multiplicities are quite large and correspond {\it a priori} to what we expect from
multifragmentation event features. At $32A$ MeV, the IMF multiplicity is
dominated by the 3-IMF channel while it goes to $5-6$ IMFs for $52A$ and 
$90A$ MeV. It is interesting to note that the IMF production saturates above $50A$
MeV (see the IMF mean values on Fig. \ref{article6}). The only change between
$52A$ and $90A$ MeV is then the IMF size, which decreases.

The correlation between
the parallel c.m. velocity and the atomic number
 (Fig. \ref{article6}, bottom row)
shows a {\it symmetric} emission of the products around the center of 
mass of the reaction (represented by the dashed line), except for the light 
charged particles ($Z<5$) where a forward-peaked contribution is visible. The energy 
thresholds of the experimental apparatus appears at backward angles in the
c.m. and truncate the velocity distribution of the backward area, showing the difficulty to
get fully detected events for this asymmetrical system with INDRA (and justify the use of the
multidimensional analysis). We will come back to this point in section 4.

\subsection{Fragmentation pattern for central collisions}

\begin{figure}[!h] 
\centering
\epsfig{file=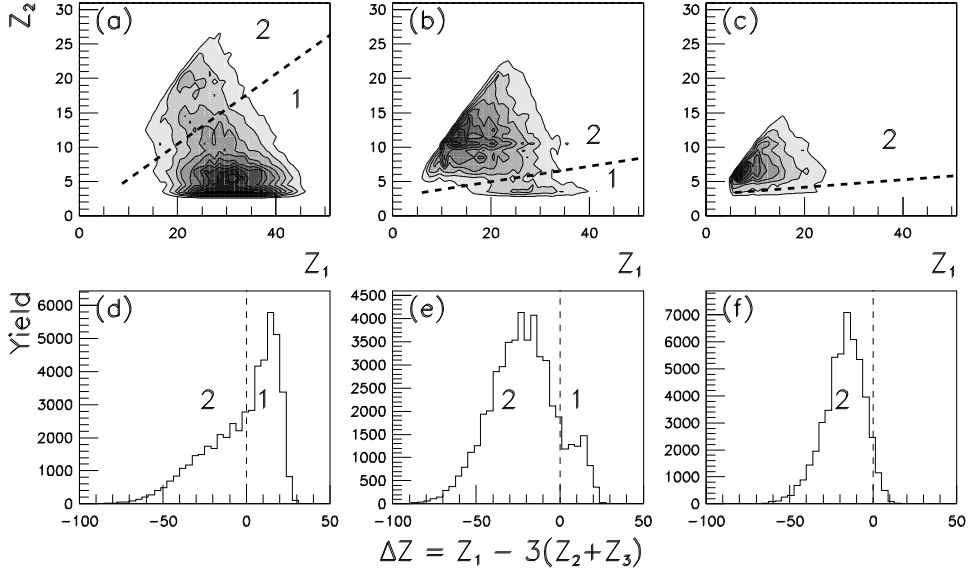,width=15cm}
\caption{$Ni + Au$ data at 32 (a,d), 52 (b,e) and 90 AMeV (c,f) (central collisions). 
Top row~: correlation between the atomic numbers of the two heaviest fragments. Bottom row~: 
charge asymmetry $\Delta Z$ defined as $\Delta Z = Z_1 - 3(Z_2+Z_3)$.}
\label{article7}
\end{figure}

Figure \ref{article7} presents the correlation between the two heaviest
fragments in the event for the $32A$ MeV(a) , $52A$ MeV(b) and $90A$ MeV(c) $Ni + Au$
system respectively. We can see clearly two zones for the first two
panels (a) and (b)~: an asymmetrical fragmentation pattern (a heavy fragment with Z around 35 
is emitted coïncidentally with a small one, Z $\approx$ 5 ), called zone $1$ and a symmetric 
fragmentation pattern (the atomic number of the two heaviest fragments is around $15-20$)
called zone $2$, while only the zone corresponding to the symmetric channel is 
present on panel (c). 

The separation between these zones depends on the incident energy because 
of the different IMF yields. This is shown on the bottom row of figure 7 
by the "$\Delta Z$" distribution (built with the the three largest atomic numbers  
 $Z_1,Z_2$ and $Z_3$) where the panel (d) and (e) exhibits a shoulder
which corresponds respectively to zone 2 (at $32A$ MeV) and 1 (at $52A$ MeV). The
definition of $\Delta Z=Z_1-3(Z_2+Z_3)$ is here arbitrary and only reflects 
the best separation we can found between the two zones.    

\begin{figure}[!h] 
\centering
\epsfig{file=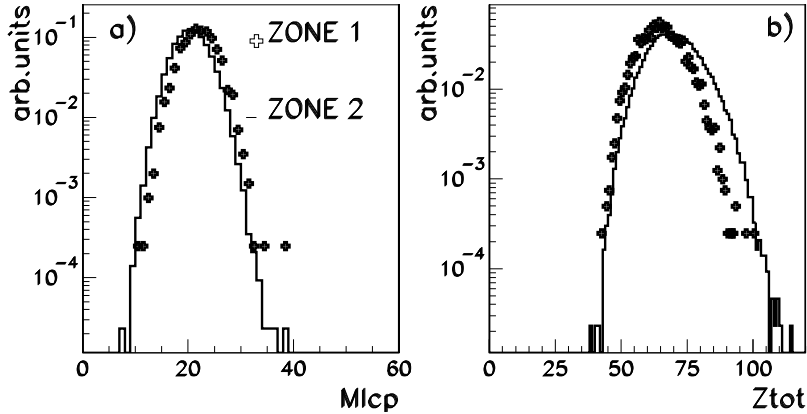,height=6cm}
\caption{$52A$ MeV $Ni + Au$ data (central collisions). LCP multiplicity (a) and 
total detected charge (b) for the zone $1$ (symbols) and $2$ (histogram).}
\label{article8}
\end{figure}

In order to check that the two fragmentation channels determined from Fig. 
\ref{article7} are not an artifact of the experimental setup (via a different detection
response of INDRA), we have plotted on Fig. \ref{article8} the LCP multiplicity and the 
total detected charge for the $2$ zones of Fig. \ref{article7} for the $52A$ MeV
$Ni+Au$ system. If we assume that 
the LCP multiplicity is related to the event dissipation, we can then conclude 
from Fig. \ref{article8}a that the events belonging to both zones are associated to 
the same class of dissipation. Indeed, the $PCA$ selection insures us that the 
selected central collision samples present homogeneous properties with respect 
to every global variable used to build the $PCA$, as already explained in 
section 3. The total detected charge distributions (b) are
 comparable, and show that the event classes have the same detection features, 
 despite their different fragmentation patterns. It demonstrates that 
 experimental biases can not account for the difference in the fragmentation pattern 
and corresponds indeed to the {\it coexistence} of two different decay channels in the
multifragmentation data. 

 In the following, we will restrain the study to the events of zone $2$ (symmetric channel), 
 which are considered as true (simultaneous) multifragmentation events, while the other 
 channel is likely related to a residue-evaporation/fission process and is rather ascribable 
 to a sequential fragment production mechanism. This statement is also supported by the 
 $\gamma$-IMF correlation analyses made for the same system at $30A$ and $45A$ MeV by the MEDEA 
 collaboration \cite{agodi} where a change from sequential to simultaneous IMF production 
 is observed for central collisions. 
 Anyway, the separation between these two
patterns is cleaner at $52A$ MeV and $90A$ MeV (where only Zone $2$ is present), while it is
difficult to define it at $32A$ MeV because of the continuity of the $\Delta Z$ distribution. Therefore we will only consider the events of Zone $2$
coming from the central collisions isolated at $52A$ and $90A$ MeV. Nevertheless, it is
important to note the coexistence of these two decay channels for the central
collisions of the $Ni+Au$ system at $32A$ and $52A$ MeV, which have been already reported 
\cite{bormio2001} 
and will be discussed in the framework of phase transition in a forthcoming paper \cite{lopez}.  

\section{Multifragmentation and equilibration}

\subsection{Comparison protocol}

In order to check the degree of equilibration of the multifragmentation
events (defined as Zone $2$ in the previous section), we need to compare the experimental 
data to the predictions of a model based upon the statistical equilibrium hypothesis.
Therefore we have used the well-known Statistical Multifragmentation Model
(SMM) of Bondorf et al \cite{smm} and applied it to the $Ni+Au$ multifragmentation events. 

\begin{figure}[!h] 
\centering
\epsfig{file=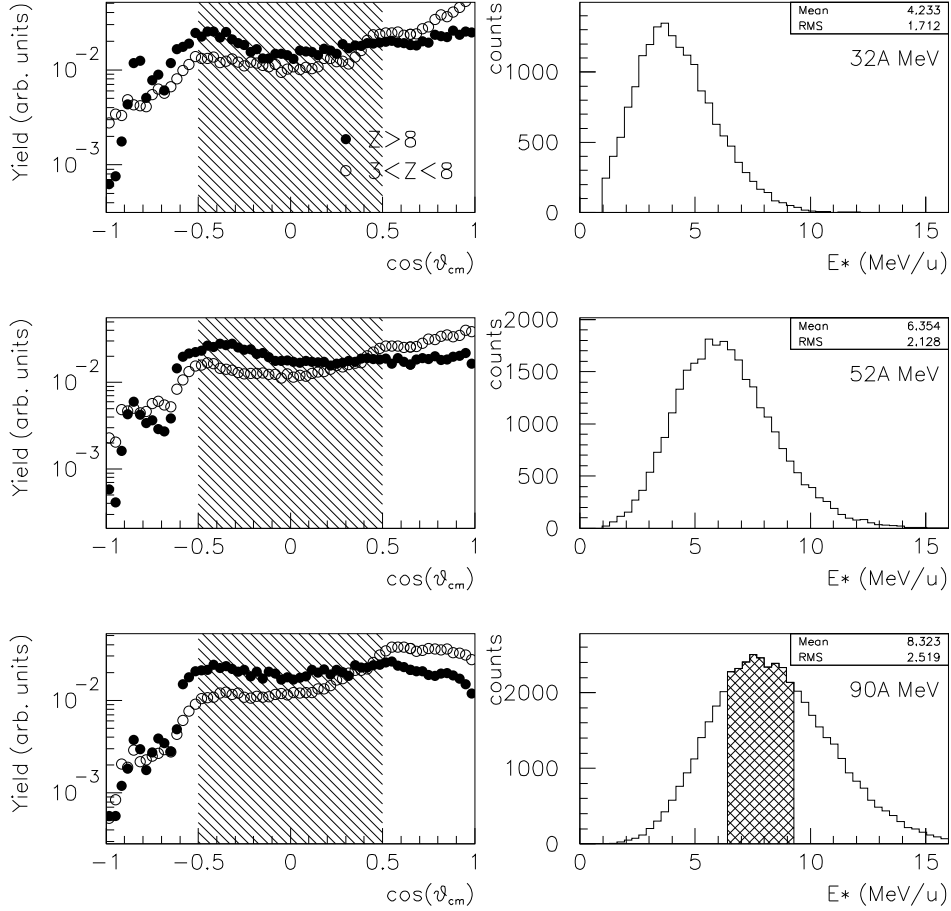,width=14cm}
\caption{Multifragmentation data. Left column~: angular distribution in the c.m. for IMFs with
$3<Z<8$ (open symbols) and $Z>8$ (solid symbols). The hatched area corresponds to the selection (see
text). Right column~: excitation energy estimated from calorimetry. The
dashed area on the bottom panel is explained in section 5.2}
\label{article9}
\end{figure}

From previous studies, it has been shown that central collisions in the Fermi energy domain 
can be affected by a pre-equilibrium component, even for IMFs \cite{colin}. 
This component is forward-peaked for the $Ni+Au$ system as illustrated on Fig.
\ref{article9}. The light fragments and particles are more affected by this effect. 
The heavy fragments have a flatter distribution
along the overall angular range despite the lack of them for the most backward angles as
already mentioned for Fig. \ref{article6}. As a matter of fact, central 
collisions emit products with low velocities in the laboratory, and the 
determination of the recoil velocity becomes rather difficult because of the energy thresholds 
imposed by the experimental setup \cite{frankland}. These two limitations have to be taken 
into account if we want to compare the data with models. Therefore, we have decided to make the 
comparisons in a restricted angular domain in the center of mass (c.m.). This domain is located 
around the transverse direction for two reasons; firstly to minimize the effect of a
bad determination of the recoil velocity (because of the lack of detection for slow fragments), 
and secondly to exclude in a large extent the pre-equilibrium component. The angular range 
intervals have been set to the same values at $52A$ and $90A$ MeV and are defined as 
$60-120^o$ (hatched area on the left column of Fig. \ref{article9}) in order to get 
reasonably flat angular distributions. In the following, all the presented data will be 
selected in this c.m. angular range.  

Figure \ref{article9} presents also the excitation energy obtained by calorimetry
\cite{leneindre} calculated in an event-by-event basis; the light charged particles
and light IMFs (up to $Z=4$) have been taken in the restricted c.m. angular range ($60-120^o$) 
and their contribution has been multiplied by 2. The evaporated neutron contribution 
has been estimated from the $N/Z$ ratio of the initial system and their kinetic energy is deduced
from the temperature determined by the kinetic equation of state as precised in ref. 
\cite{dagostino2}. The distributions are quite broad and do not only reflect the fluctuations in
the energy deposition but also the uncertainties brought by the experimental determination 
(neutrons, pre-equilibrium component not completely removed, detection). Nevertheless, we can note 
the mean values ($4.2$,$6.4$,$8.3A$ MeV for respectively $32$,$52$,$90A$ MeV) and use 
them as starting points for the following model comparisons. 

Table \ref{edisp} gives the available c.m. energy (total and per nucleon) for the studied systems
and by comparing with the extracted excitation energy values we have an estimation of the
percentage of deposited energy as a function of the asymmetry of the system (when comparing $Ni+Au$
to $Xe+Sn$) or the incident energy ($Ni+Au$). The value for the $Xe+Sn$ system ($9.3A$ MeV) 
has been obtained by using the same prescription \cite{bellaize}.

\begin{table}[!h]
\begin{center}
\begin{tabular}{ccccc}\hline
 System & $E_{inc}$ ($A$MeV) & $E_{cm}$ (MeV) & $E_{cm}$ (MeV/n) & $E^{*}/E_{cm}$ \\ \cline{1-5}  
 $^{58}Ni+^{197}Au$ & 32 & 1430 & 5.6 & 0.81 \\
 $^{58}Ni+^{197}Au$ & 52 & 2320 & 9.1 & 0.70 \\
 $^{58}Ni+^{197}Au$ & 90 & 4000 & 15.7 & 0.53 \\  
 $^{119}Xe+^{nat}Sn$ & 50 & 3080 & 12.4 & 0.73  \\ \cline{1-5}
\end{tabular}
\vskip 0.2cm
\caption{ Incident energy $E_{inc}$, available c.m. energy $E_{cm}$ and ratio of estimated 
deposited energy $E^{*}/E_{cm}$ for the studied systems.}
\label{edisp}
\end{center}
\end{table}

The ratio between the deposited energy (estimated by calorimetry) and the c.m. available energy
decreases as a function of the incident energy from $0.8$ at $32A$ MeV to $0.5$ at $90A$ MeV for
the $Ni+Au$ system; it shows the ever increasing part of the preequilibrium component when 
going from Fermi energy toward the relativistic regime. Moreover, we do
not observe any dependence of the ratio ($\approx 0.7$) as a function of the mass asymmetry of the 
entrance channel for a given incident energy (here $50A$ MeV). Thus, no transparency effects are 
evidenced at that incident energy.
      
\subsection{Partitions}

\begin{figure}[!h] 
\centering
\epsfig{file=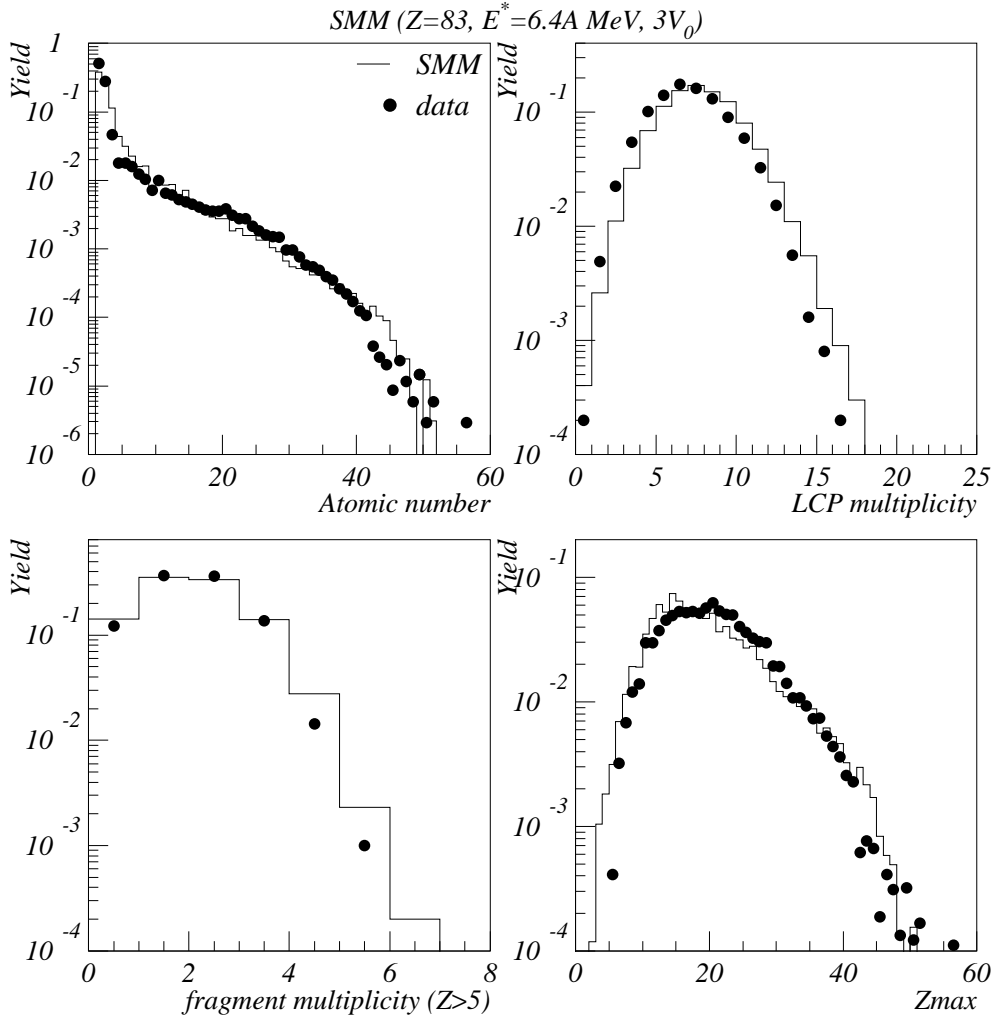,width=14cm}
\caption{Atomic number (a) for the $52A$ MeV $Ni+Au$ data (symbols) and SMM
(histogram), LCP multiplicity ($Z\leq2$) (b), Fragment multiplicity ($Z\geq6$) (c) and atomic number 
distributions of the heaviest fragment (d). These comparisons are made in a restricted 
angular domain in the center of mass ($60-120^{o}$). All spectra are
normalized to the number of events.}
\label{article10}
\end{figure}

Figure \ref{article10} presents the results obtained for the comparison between the
experimental data (here $52A$ MeV $Ni+Au$) and the simulation (SMM) for the best set
of input parameters we have found; it corresponds to a source size of $A=215$
($Z=86$) with an thermal excitation energy of $E^{\ast}/A=6.5 MeV$. This latter
value is close to the one obtained experimentally by calorimetry 
($<E^{\ast}/A>=6.4$ MeV) \cite{bellaize} and fig. \cite{article9} right. Several attempts have been made in order to 
determine the optimal freeze-out volume in the model but due to the comparison in a rather 
small angular range, no sensitivity to this parameter has been found and
we have taken the usual value of $3V_{0}$ for the freeze-out configuration.

The data (symbols) for the $52A$MeV $Ni+Au$ system are fairly reproduced by
the model (histogram) as far as partitions are concerned. The atomic number
distribution (Fig. \ref{article10}a) exhibits some discrepancies for charges
between $20$ and $35$ but the global trend of the distribution is
reproduced. It has to be mentioned that the spectra of Fig. \ref{article10}a 
are normalized to the number of events. The LCP multiplicity (Fig. \ref{article10}b) 
and the fragment multiplicity ($Z>5$, Fig. \ref{article10}c) are slightly overestimated 
by the model while the charge of the heaviest fragment (Fig. \ref{article10}d) gets the same
agreement as the inclusive charge distribution. The events generated by SMM
have been passed through a software replica of the INDRA array and the
comparisons are made into a restricted angular range domain in the center of
mass (here between $60$ and $120$ degrees) in order to minimize the threshold
biases and pre-equilibrium effects \cite{bellaize,leneindre}, as already explained 
in section 4.1.

Some attempts have been made in order to get a more quantitative comparison between the data and 
the model predictions, such as it is preconized in \cite{desesquelles_backtracing} by the {\it backtracing}
 method. The results show that the confidence intervals where the data are reasonably reproduced are 
$0.5A$ MeV for the excitation energy and $5$ charge units for the source size. This agreement has been
numerically evaluated by the calculation of a $\chi^2$ upon the $4$ distributions presented on Fig.
\ref{article10}. The confidence intervals given here refer to minimal $\chi^2$ values which 
do not differ from more than $10\%$, and the values reported for the model corresponds 
to the central points of these intervals.

\begin{figure}[h] 
\centering
\epsfig{file=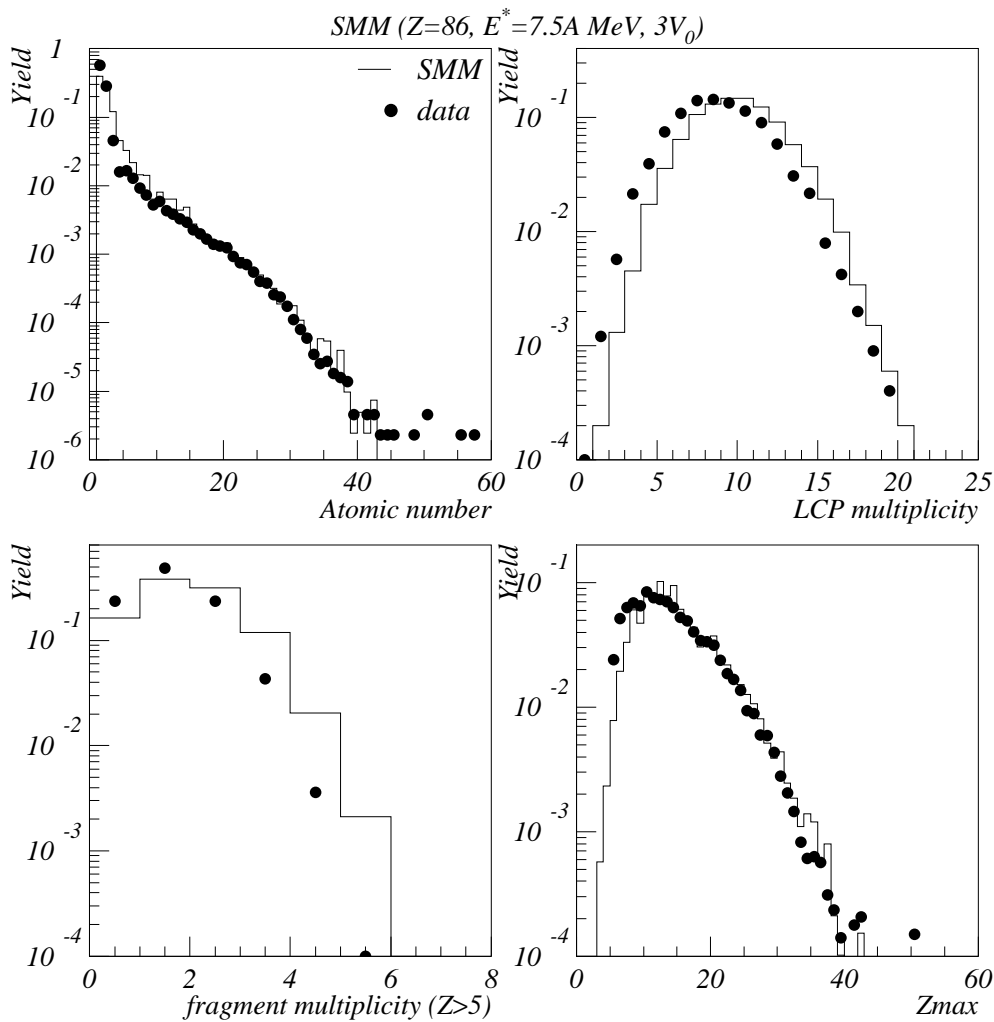,width=14cm}
\caption{Atomic number distribution (a) for the $90A$ MeV $Ni+Au$ data 
(symbols) and SMM (histogram), LCP multiplicity ($Z\leq2$) (b), 
Fragment multiplicity ($Z\geq6$) (c) and atomic number 
distributions of the heaviest fragment (d). These comparisons are made in a restricted angular domain in
the c.m. ($60-120^{o}$). All spectra are normalized to the number of events.}
\label{article11}
\end{figure}

The same comparison has been made for the $90A$MeV $Ni+Au$ system and
is presented on Fig. \ref{article11}. The input parameters of SMM are 
now $A=215$ ($Z=86 \pm 5$) and $E^{*}/A = 7.5 \pm 0.5$ MeV. The agreement is 
satisfactory between the model and the data, showing then that the fragment 
partitions are also consistent with a statistical exploration of the available 
phase space as it is suggested in the SMM model.

\subsection{Kinetic properties}

\begin{figure}[h] 
\centering
\epsfig{file=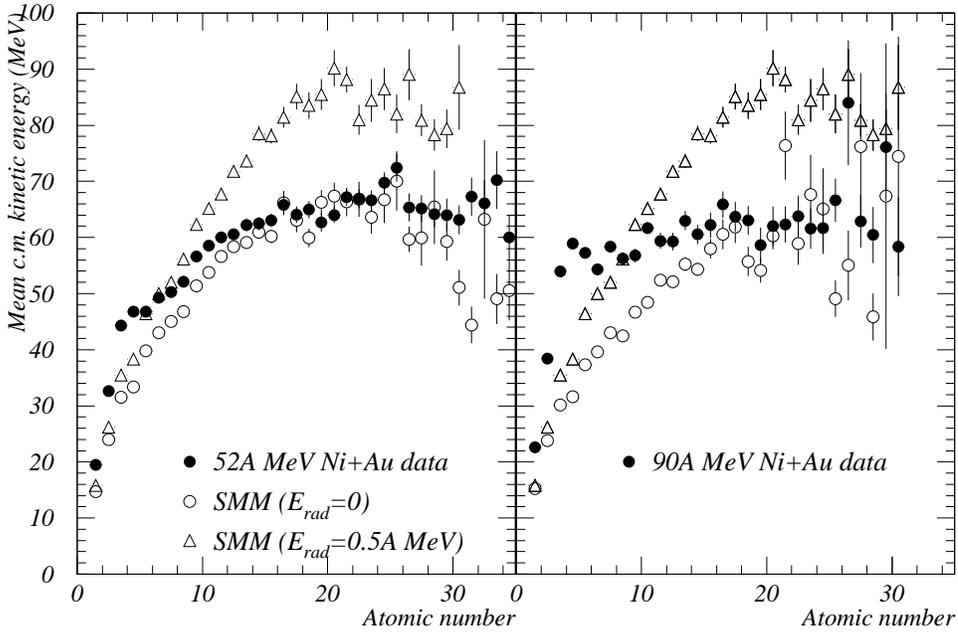,width=14cm}
\caption{Mean c.m. kinetic energy versus atomic number for the multifragmentation events of the 
$Ni+Au$ system at $52A$ MeV (left) and $90A$ MeV (right) (solid circles), 
SMM calculations without radial flow (open circles) and SMM calculations with $0.5A$ MeV of radial flow 
(triangles).}
\label{article12}
\end{figure}

Figure \ref{article12} shows the correlation between the mean kinetic energy
(in the c.m.) and the atomic number for the data (solid circles) and 
the model (open symbols). 
An agreement is accounted for the heaviest atomic number ($Z>15$) as displayed on
Fig. \ref{article12} (left panel for $52A$MeV and right for $90A$MeV). 
However, the model underestimates in a
large extent the mean kinetic energies for light charged particles and IMFs
($Z<10$). If we assume an extra collective energy (radial flow)
of $0.5A$MeV in the model (open triangles) , then the mean kinetic
energies are slightly better reproduced for $Z<10$ but the agreement fails totally
for the heavier products. 

We can therefore conclude that no radial flow has to be put in SMM calculations to reproduce
the kinetic properties of fragments while strong out-of-equilibrium features appear for light
charged particles and light IMFs ($Z<10$). This result is somehow consistent with the conclusions 
made from the $Ar+Au$ analysis in the same incident energy range \cite{desouza93} 
where is evidenced an expansion energy component; indeed, in this
case, only kinetic energy distribution for light IMFs ($Z<10$) have been analysed and give the same 
mean values as our analysis. This important point will be furthermore developed in the next 
section by the direct comparison between experimental data.

\section{Experimental comparisons}

At this point, we have shown that the multifragmentation events are compatible
with a scenario of statistical partitions as predicted by SMM. We will now go
further by systematically comparing data coming from systems 
with different entrance channels. To do so, we will use the same
$^{58}Ni+^{197}Au$ 
multifragmentation events as before (at $52A$ and $90A$ MeV) but also a new one
 obtained by the same selection method ($PCA$) onto the $50A$ MeV
$^{129}Xe+^{nat}Sn$ system \cite{bellaize,leneindre}. 
We will have then 2 systems with different mass asymmetries in
the entrance channel ($Xe+Sn$, $Ni+Au$), and two incident velocities 
($50A/52A$, $90A$MeV) and will present in the following 
the cross-comparisons between them.  
   
\subsection{Same incident energy}

\begin{figure}[!h] 
\centering
\epsfig{file=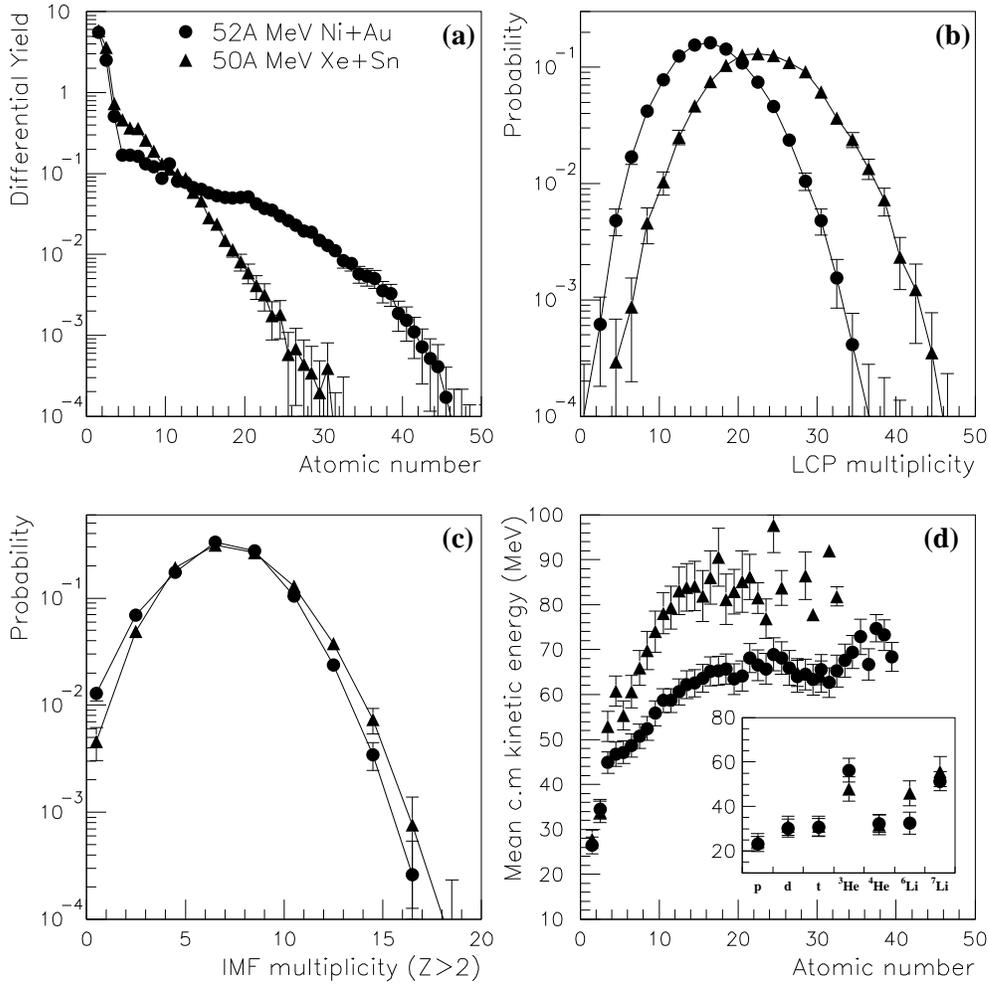,width=14cm}
\caption{Comparison between $52A$ MeV $Ni+Au$ and $50A$ 
MeV $Xe+Sn$ multifragmentation data. Atomic number (a), LCP multiplicity (b) , IMF
multiplicity (c) and correlation between the mean c.m. 
kinetic energy and the atomic number (d) ($Ni+Au$ : circles, $Xe+Sn$ : triangles). 
These comparisons are made in a restricted angular domain in the c.m. 
($60^{o}-120^{o}$). All spectra are normalized to the number of events. The LCP and IMF
multiplicities have been multiplied by two in order to reflect the integrated values.}
\label{article13}
\end{figure}

The comparison between two systems ($52A$ MeV $Ni+Au$ and $50A$ MeV $Xe+Sn$) 
of nearly the same incident velocity and the same total size ($250$ nucleons) 
is presented on Fig. \ref{article13}. The
data correspond to multifragmentation samples (using the $PCA$
selection), and have been characterized by almost the same source size ($<Z>=85$
for the $50A$ MeV $Xe+Sn$ system) but different thermal excitation energies 
($7.5A$ MeV compared to $6.4A$ MeV for $Ni+Au$ once substracted the radial energy component) 
\cite{bellaize,leneindre}. We see clearly that the charge distribution (\ref{article13}a) or
 the LCP (\ref{article13}b) multiplicities (multiplied by 2 in order to get integrated values) 
 extend higher for the $50A$MeV $Xe+Sn$ system as expected by
the thermal excitation energy difference of about $1A$ MeV. Conversely, the IMF
multiplicities (\ref{article13}b) are similar but only reflect the counterbalance between the
higher IMF ($3<Z<15$) and the smaller heavy fragment ($Z>15$) yields for the $Xe+Sn$ system.  

Figure \ref{article13}d shows the mean c.m. kinetic energies as a function of the
atomic number. We see here that the light charged particles (up to Lithium) are superimposed 
(insert of Fig.\ref{article13} d) while a major discrepancy appears for heavier fragments. 
Indeed, in the $Xe+Sn$ system case, a radial energy component has to be 
invoked as shown in \cite{marie,leneindre}. This is not the 
case for the $Ni+Au$ system for which one does not observe the linear increase of the mean kinetic 
energy as a function of the fragment charge, as expected if there is a radial flow 
component. 

\subsection{Same thermal excitation energy}

\begin{figure}[!h] 
\centering
\epsfig{file=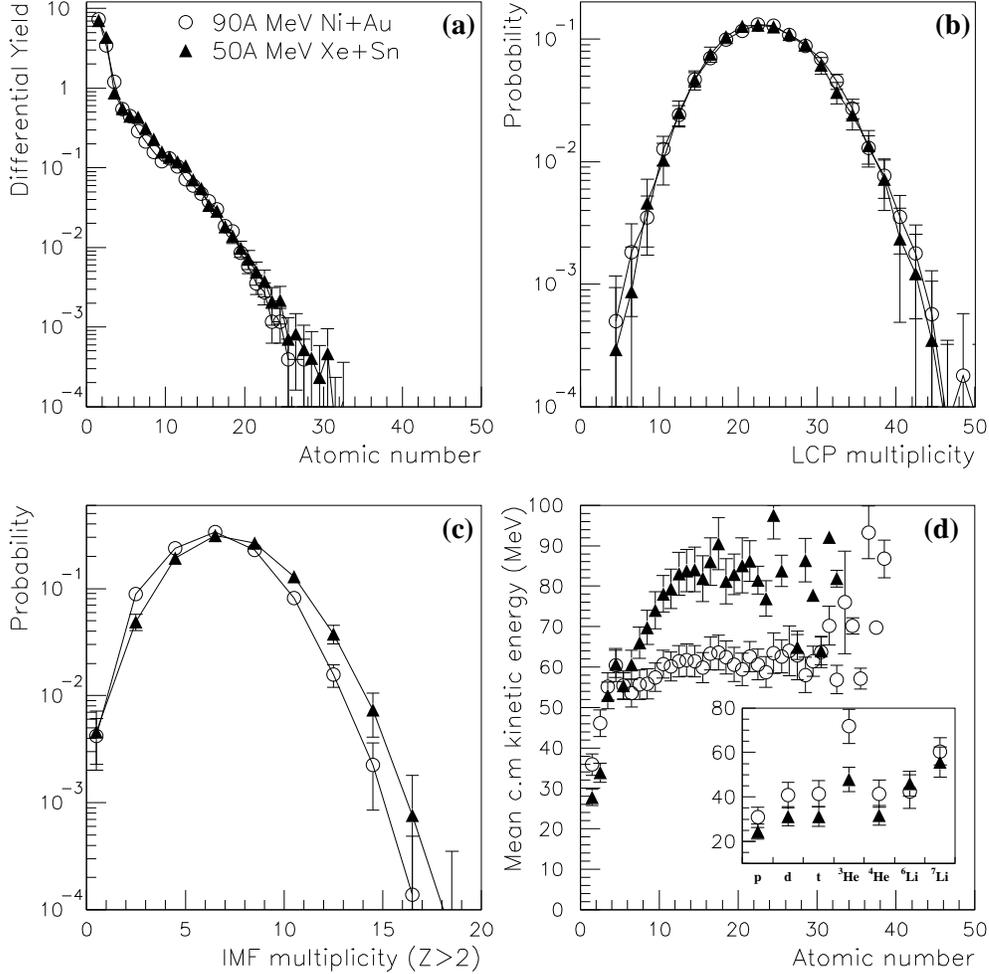,width=14cm}
\caption{Comparison between $90A$ MeV $Ni+Au$ and $50A$ MeV $Xe+Sn$ 
multifragmentation data. 
Atomic number distribution (a), LCP multiplicity (b), IMF
multiplicity (c) and correlation
between the mean c.m. kinetic energy and the atomic number (d) ($Ni+Au$ : open
circles, $Xe+Sn$ : solid triangles). These comparisons are made in a restricted angular domain in
the c.m. ($60^{o}-120^{o}$). The data have been gated around the following excitation
energy interval : $6 - 9A$ MeV. All spectra are normalized to the number 
of events. The LCP and IMF multiplicities have been multiplied by two in order 
to reflect the integrated values.}
\label{article14}
\end{figure}

The comparison is now made between the $50A$ MeV $Xe+Sn$ and $90A$ MeV
$Ni+Au$ systems (Fig. \ref{article14}). Here, the mean thermal excitation energy 
components, determined experimentally and by comparison with the SMM model, are nearly 
equivalent \cite{bellaize} and are around $7.5-8A$ MeV. Nevertheless, we decide to present
the data with a condition on the (thermal) excitation energy
because of the broadness of the excitation energy distribution (more than $7A$
MeV for the $Ni+Au$ system); to do so, we have gated around the most probable 
excitation energy value (see the dashed area on Fig. \ref{article9}, bottom panel of 
the right column), and it corresponds here to $7.7 \pm 1.5A$ MeV. In this way, 
the excitation energy domain is similar for both systems.
 
From Figure \ref{article14}a, we see that the atomic number 
distributions are nearly identical, and that the LCP and IMF multiplicities 
(\ref{article14}b and \ref{article14}c) 
are comparable, showing that the fragmentation pattern is 
equivalent between these two systems, and is governed by the
thermal excitation energy stored into the system. This result means that the partitions are 
similar whatever the radial 
flow is. Figure \ref{article14}d, displaying 
the mean c.m. kinetic energy as a 
function of the charge shows a disagreement for fragments ($Z>3$), while the
situation is different for light charged particles ($Z<4$) as seen in the insert. 
For these particles, the mean kinetic energies are {\it higher} for the $Ni+Au$ system, as
expected for pre-equilibrium emission, mainly sensitive to the incident energy.
The large disagreement observed for fragments is attributed to the 
existence of an additionnal collective energy in the $Xe+Sn$ system of 
$2A$MeV \cite{bouriquet,leneindre}. Putting together these obsevations, 
it demonstrates the decoupling between the thermal 
and the radial flow components of the excitation energy. This statement seems to 
be valid in the limit of a small 
ratio between the collective and the thermal component because large 
deviations to
the decoupling have been found at higher incident energies \cite{hsi,lauret}.

\subsection{Excitation function of the $Ni+Au$ system}

\begin{figure}[h]
\centering
\epsfig{file=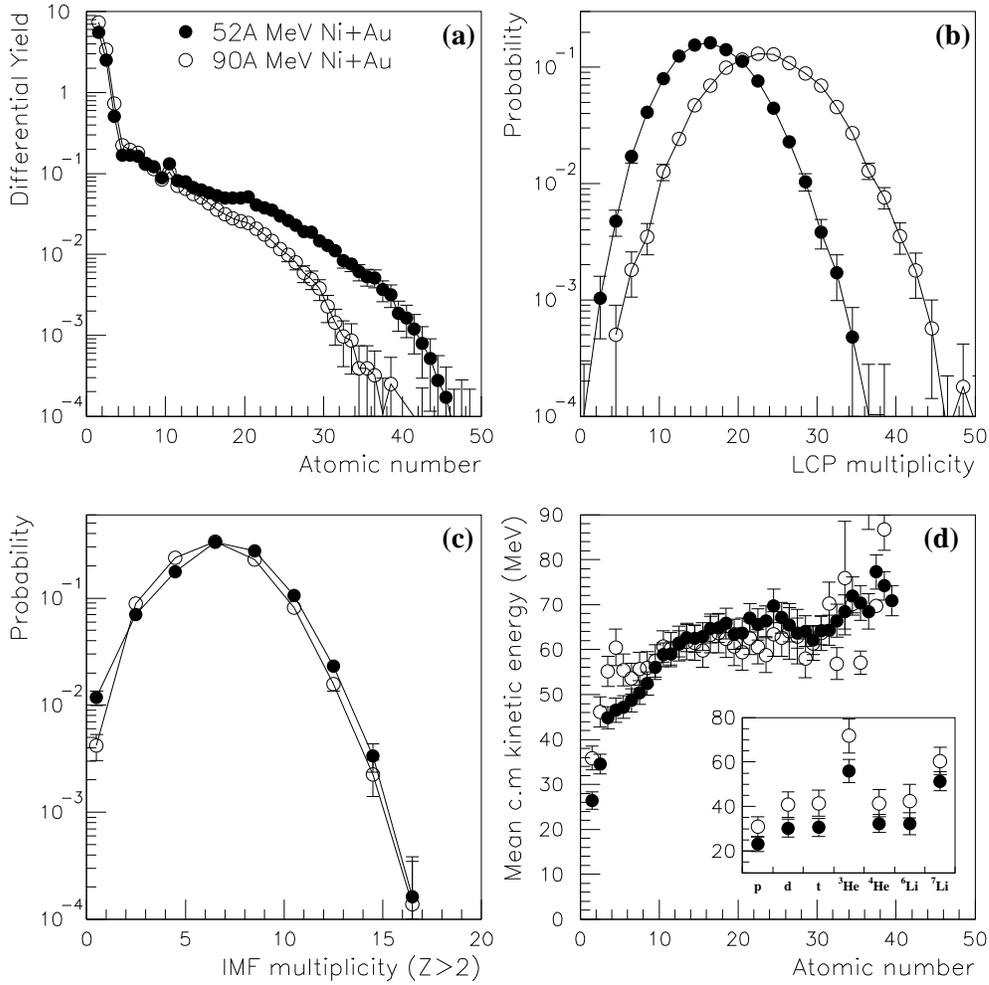,width=14cm}
\caption{Comparison between $52A$ MeV and $90A$ MeV $Ni+Au$ multifragmentation data. 
Atomic number (a), LCP multiplicity (b),
IMF multiplicity (c) and correlation
between the mean c.m. kinetic energy and the atomic number (d) ($52A$ MeV : solid
circles, $90A$ MeV : open circles). These comparisons 
are made in a restricted angular domain in the c.m. ($60^{o}-120^{o}$. 
All spectra are normalized to the number of events. The LCP and IMF
multiplicities have been multiplied by two in order to reflect the integrated values.}
\label{article15}
\end{figure}

At last, comparisons for the same system ($Ni+Au$) at different incident
energies ($52A$ and $90A$ MeV) are presented in Fig. \ref{article15}. 
The fragmentation pattern is different as illustrated on Fig. \ref{article15} 
(atomic number 
(a) and LCP multiplicity (b) distributions). Figure \ref{article15}c 
presents the IMF multiplicities which are surprisingly equivalent. It means that 
the maximum number of fragments has been already reached for the central collisions of 
the $Ni+Au$ system between $52A$ and $90A$ MeV, while the LCP multiplicity
continues to increase (Fig. \ref{article15}b); in fact, this is the 
size of the fragments which decreases when going from $52A$ to $90A$ MeV 
but not their yield. Systematic studies of IMF production for symmetric 
systems \cite{peaslee,reisdorf,sisan} have shown that a maximum
was reached at incident energies increasing with the mass of the system and was
located around $100A$ MeV, while it has been found at smaller incident energies 
- around $50A$ MeV - for the $Xe+Sn$ systems measured by the INDRA collaboration \cite{rab99}. 
For the (asymmetrical) $Ni+Au$ system, we also come to the same conclusions (maximum around $50A$
MeV) and it can therefore be related to a common behavior of fused systems \cite{frankland}.
  
The kinetic properties, displayed by Fig. \ref{article15}d, are 
also similar, and substantiate the inexistence of any radial flow energy 
in the data. The light charged particles, shown in the insert, are systematically higher
for the $90A$ MeV system (open circles) as expected both from 
pre-equilibrium and thermal effects.  

At this point, we must recall that the source sizes, determined both from SMM
comparisons and experimentally, are comparable \cite{bellaize};
 the difference in the
{\it thermal} component of the excitation energy between the two systems is roughly  
compensated by the difference between the Coulomb energies at the freeze-out stage, 
leading to the same kinetic energy profiles. This suggests that we need to compare both 
{\it static} and {\it kinetic} observables to get relevant information 
about the characteristics of fused systems formed in central collisions.

\section{Conclusions}

We have presented a new technique in order to carefully isolate 
the most violent collisions by using a 
multidimensional analysis named the Principal component Analysis ($PCA$). 
We have shown that this method can be 
applied for all studied systems over a large domain of incident energy 
(i.e. between $32A$ and $90A$MeV) and is perfectly suited for $4\pi$ devices. 
This provides a powerful tool to select 
event classes in the domain of heavy-ion induced reactions. The method have
then been used to select multifragmentation data among the most dissipative collisions.
The selected samples have been found to be highly dissipative and to present two
different fragmentation patterns (at $32A$ and $52A$ MeV) which coexists at 
the same excitation energy. The first one is
associated to residue-evaporation or fission-evaporation mechanism and is
characterized by a big residue ($Z \approx 35$)  while the second mechanism is
associated to a more copious IMF production and can be related to the
multifragmentation of a single source.   

We have analysed the multifragmentation events from the $Ni+Au$ system at $52A$ and 
$90A$MeV. From the comparisons with the SMM model and between experimental data, we have 
evidenced that fragment partitions are mainly governed by the thermal 
part of the excitation energy. Conversely, for light charged particles and light IMFs, we
have seen that their kinetic properties are closely related to the incident
energy and mass asymmetry, enlightening thus the different emission time scales between 
these products and the heavier ones, showing then that only a full treatment of the
dynamics of the collision can provide relevant comparison with data. 
No evidence for a sizeable radial flow energy 
has been found for the $Ni+Au$ system between $52A$ and $90A$MeV 
nor any dependence as a function of the incident energy. The 
fragmentation pattern are very similar for the $90A$ MeV $Ni+Au$ and
 the $50A$ MeV $Xe+Sn$ systems while the kinetic
features are found to be completely different. This fact shows clearly that a
decoupling between the thermal and the radial flow part of the excitation
energy is achieved in this incident energy domain where the radial component is quite
small as compared to the thermal one. 

By comparing the observed difference in the energy deposition (thermal + expansion), we find that 
it is maximal for the symmetric system (conversely the pre-equilibrium component is
minimised) and constitutes then the most efficient way to dissipate energy into nuclear systems. 
As a final conclusion, we would say that 
nuclei cannot be produced with {\it thermal} excitation energies greater 
than the value given by the average binding energy ($\approx 8A$ MeV). This means 
that the extra energy dissipated into the system ($50A$ MeV $Xe+Sn$ case) is then 
entirely converted into collective degrees of freedom such as the radial flow 
of particles.

\end{document}